\documentclass[final,3p,times,twocolumn]{elsarticle}

\usepackage{amssymb}
\usepackage{subfigure}
\usepackage{color}

\usepackage{lineno}

\journal{Nuclear Physics A}

\begin{document}

\begin{frontmatter}



\title{Evaluation of Kyoto's Event-Driven X-ray Astronomical \\ SOI Pixel Sensor with a Large Imaging Area}


\author[kyoto]{Hideki Hayashi}
\ead{hayashi@cr.scphys.kyoto-u.ac.jp}
\author[kyoto]{Takeshi Go Tsuru}
\ead{tsuru@cr.scphys.kyoto-u.ac.jp}
\author[kyoto]{Takaaki Tanaka} \author[kyoto]{Hiroyuki Uchida} \author[kyoto]{Hideaki Matsumura} 
\author[kyoto]{Katsuhiro Tachibana} \author[kyoto]{Sodai Harada}
\author[miyazaki]{Ayaki Takeda} \author[miyazaki]{Koji Mori} \author[miyazaki]{Yusuke Nishioka} \author[miyazaki]{Nobuaki Takebayashi}
\author[miyazaki]{Shoma Yokoyama} \author[miyazaki]{Kohei Fukuda}
\author[kek]{Yasuo Arai} \author[kek]{Ikuo Kurachi}
\author[shizuoka]{Shoji Kawahito} \author[shizuoka]{Keiichiro Kagawa} \author[shizuoka]{Keita Yasutomi} \author[shizuoka]{Sumeet Shrestha} \author[shizuoka]{Syunta Nakanishi}
\author[okinawa]{Hiroki Kamehama}
\author[tokyorika]{Takayoshi Kohmura} \author[tokyorika]{Kouichi Hagino} \author[tokyorika]{Kousuke Negishi} \author[tokyorika]{Kenji Oono} \author[tokyorika]{Keigo Yarita}

\address[kyoto]{Department of Physics, Graduate School of Science, Kyoto University, Kitashirakawa Oiwake-cho, Sakyo-ku, Kyoto 606-8502, Japan}
\address[miyazaki]{Department of Applied Physics, Faculty of Engineering, University of Miyazaki,1-1 Gakuen Kibana-dai Nishi, Miyazaki 889-2192, Japan}
\address[kek]{Institute of Particle and Nuclear Studies, High Energy Accelerator Research Org., KEK, 1-1 Oho, Tsukuba 305-0801, Japan}
\address[shizuoka]{Research Institute of Electronics, Shizuoka University, Johoku 3-5-1, Naka-ku, Hamamatsuu, Shizuoka 432-8011, Japan}
\address[okinawa]{National Institute of Technology, Okinawa College, Henoko 905, Nago-shi, Okinawa 905-2192, Japan}
\address[tokyorika]{Department of Physics, Faculty of Science and Technology, Tokyo University of Science, 2641 Yamazaki, Noda, Chiba 278-8510, Japan}

\begin{abstract}

We have been developing monolithic active pixel sensors, named ``XRPIX'', based on the silicon-on-insulator (SOI) pixel technology for future X-ray astronomy satellites.
XRPIX has the function of event trigger and hit address outputs.
This function allows us to read out analog signals only of hit pixels on trigger timing, which is referred to as the event-driven readout mode.
Recently, we processed ``XRPIX5b'' with the largest imaging area of 21.9~mm $\times$ 13.8~mm in the XRPIX series.
X-ray spectra are successfully obtained from all the pixels, and the readout noise is 46~e$^-$~(rms) in the frame readout mode.
The gain variation was measured to be 1.2\%~(FWHM) among the pixels. 
We successfully obtain the X-ray image in the event-driven readout mode.

\end{abstract}


\end{frontmatter}


\section{Introduction}
\label{Introduction} 

X-ray charge-coupled devices (CCDs) are standard imaging spectrometers widely used for focal plane detectors in modern X-ray astronomy satellites \cite{G.P.Garmire, L.Struder, M.J.L.Turner, K.Koyama, N.Meidinger, H.Tsunemi, T.Tanaka}. 
CCDs have small pixel sizes of $\sim$~20~$\mu$m~$\times$~20~$\mu$m
and good energy resolution of $\sim$ 130~eV in full width at half maximum (FWHM) at 6~keV.
However, CCDs suffer from limited time resolution of $\sim$sec for the conventional MOS-CCDs and $\sim$ 10~msec for the pn-CCDs.
Therefore, we have been developing ``XRPIX'' with a high time resolution  better than 10~$\mu$sec for future X-ray astronomy satellites \cite{T.G.Tsuru, T.G.Tsuru2018}. 

XRPIX is a monolithic active pixel sensor based on the silicon-on-insulator (SOI) pixel technology \cite{Y.Arai}.
XRPIX contains a comparator circuit in each pixel for hit trigger (timing) and two-dimensional hit-pattern (position) outputs.
The function allows us readout of pulse heights only from hit pixels on trigger output timing, which is referred to as the event-driven readout mode.
Thus, XRPIX offers a high time resolution better than 10~$\mu$sec and a high throughput reaching $>$~1~kHz in addition to the imaging and spectroscopic capabilities comparable to those of CCDs.
Taking advantage of the good time resolution,
we can significantly reduce non-X-ray backgrounds especially above 10~keV due to interactions of cosmic rays by introducing an anti-coincidence technique. 
In our previous studies, we successfully demonstrated the X-ray detection in the event-driven readout mode \cite{A.Takeda2013}.

Recently, we processed ``XRPIX5b'' with the largest imaging area in the XRPIX series.
In this paper, we report the first evaluation results of XRPIX5b.
We describe XRPIX5b in Section \ref{Device description of XRPIX5b}.
In Section \ref{Spectral performance}, we describe the depletion depth and the X-ray spectrum.
The conversion gain uniformity is shown in Section \ref{Gain uniformity}.
We evaluate the event-driven readout in Section \ref{Event-driven readout},
and summarize the evaluation results in Section \ref{Summary}. 

\section{Device description of XRPIX5b}
\label{Device description of XRPIX5b} 

\begin{figure}[h]
\centering
\includegraphics[width=7.5cm]{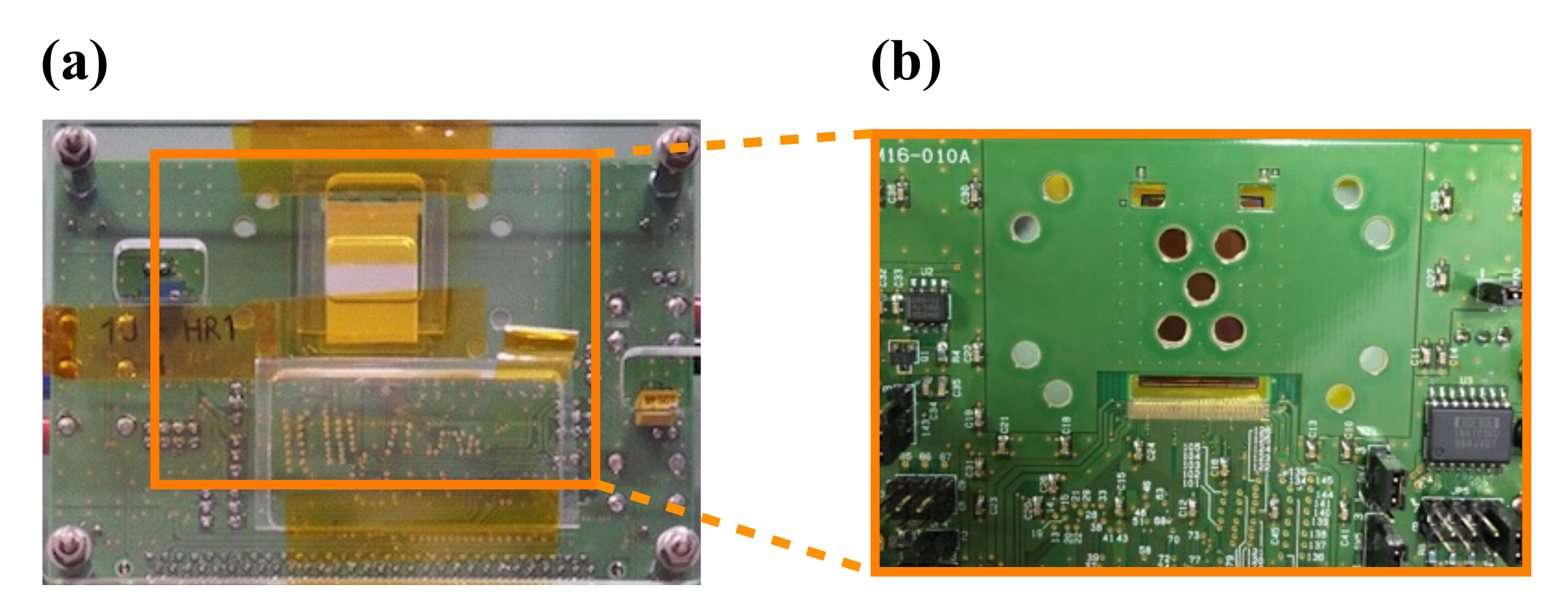}
\caption{Photographs of XRPIX5b and the readout board as seen from the sensor layer side (a) and from the circuit layer side (b). An instrumentation amplifier is also mounted on the circuit layer as a second stage amplifier.}
\label{fig:Picture_XR5b}
\end{figure}

\begin{figure}[h]
\centering
\includegraphics[width=7.5cm]{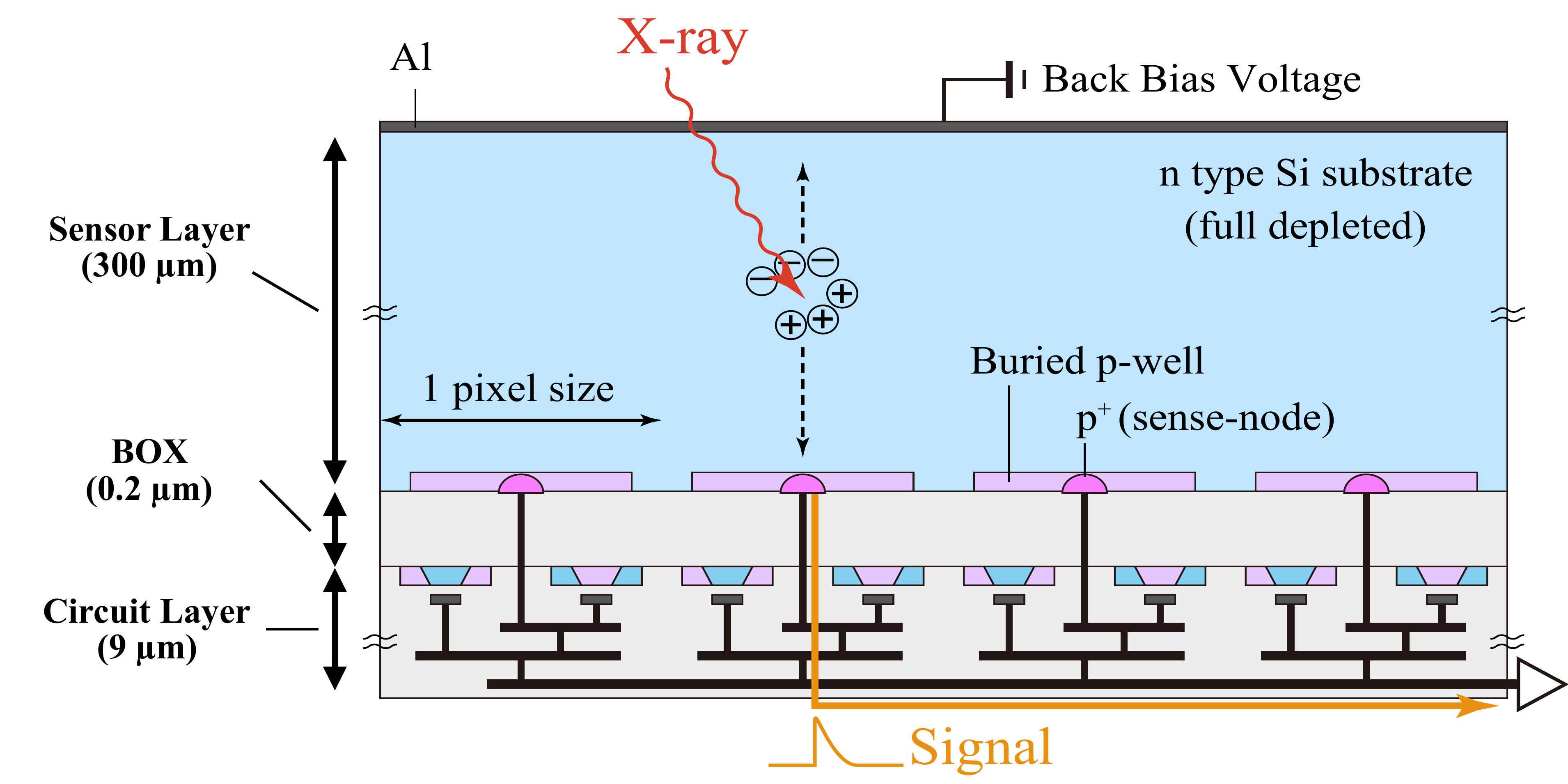}
\caption{Cross-sectional view of XRPIX5b.}
\label{fig:CrossSection}
\end{figure}

XRPIX5b is the first device in the XRPIX series having a large imaging area usable for a focal plane detector in X-ray astronomy satellites.
We used the 0.2~$\mu$m fully depleted SOI CMOS process by Lapis semiconductor Co., Ltd. 
Fig.~\ref{fig:Picture_XR5b} shows photographs of XRPIX5b and the readout board.
The imaging area, the format and the pixel size are 21.9~mm~$\times$~13.8~mm, 608~$\times$~384~pixels and 36~$\mu$m~$\times$~36~$\mu$m, respectively.
XRPIX5b was designed based on ``XRPIX3b'' \cite{A.Takeda2015}, which has a small imaging area of 1~mm~$\times$~1~mm and almost the same pixel circuit as XRPIX5b.
Fig.~\ref{fig:CrossSection} shows a cross-sectional view of XRPIX5b.
XRPIX5b consists of the following three layers:
a high-resistivity depleted silicon layer for X-ray detection (sensor layer), a CMOS circuit layer for signal processing (circuit layer), and a buried oxide (BOX) layer for insulation between the two layers.
A buried p-well (BPW) is implanted around the sense node with a size of 21~$\mu$m~$\times$~21~$\mu$m for efficient signal charge collection and to suppress the back-gate effect \cite{Y.Arai}.
The sensor layer is fabricated on n-type float zone wafer with a thickness of 300~$\mu$m. 

\section{Spectral performance}
\label{Spectral performance}

\begin{figure}[h]
\centering
\includegraphics[width=7.8cm]{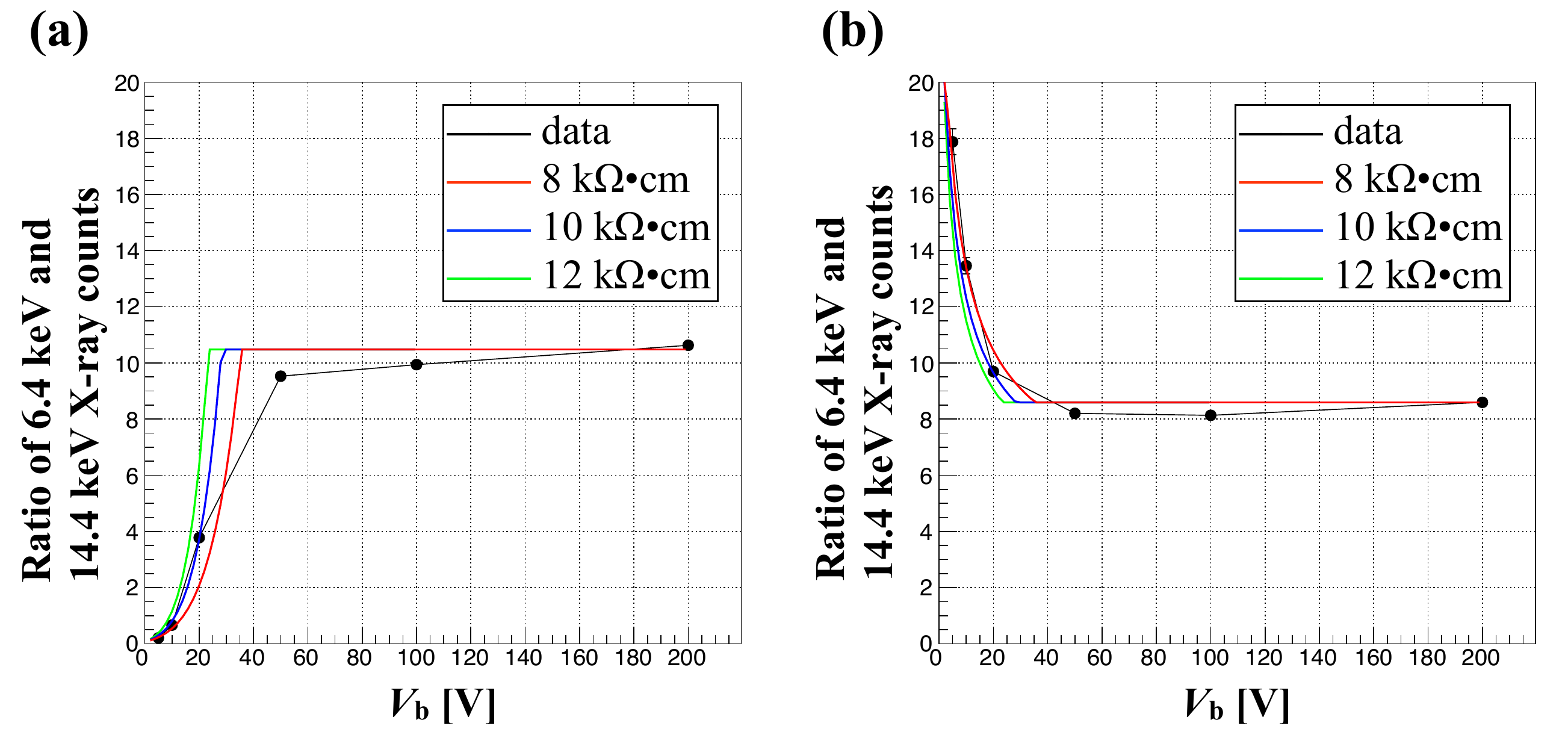}
\caption{(a) Ratios of 6.4~keV and 14.4~keV X-ray counts obtained with backside illumination as a function of $V_{\rm b}$.
(b) Same as (a) but with frontside illumination.}
\label{fig:Ratio_FI_BI}
\end{figure}

In order to estimate the depletion depth, we obtained $^{57}$Co X-ray spectra and measured ratios of 6.4~keV and 14.4~keV X-ray counts with several back-bias voltages ($V_{\rm b}$) from 5~V to 200~V.
In the measurements, we cooled XRPIX5b to $-60$~${}^\circ \mathrm{C}$ in order to reduce leakage current. 
We use all the X-ray events classified as single-pixel, double-pixel, triple-pixel, and quad-pixel events. 
The details of the readout sequence and spectral analyses are presented by Ryu et al. (2011) and Nakashima et al. (2012)\cite{Ryu, Nakashima2012}.
Fig.~\ref{fig:Ratio_FI_BI} (a) and (b) shows the ratios obtained with backside illumination (BI) and frontside illumination (FI) as a function of $V_{\rm b}$,
indicating that the ratios are almost unchanged at $V_{\rm b}$~$\gtrsim$~50~V.
This suggests that the quantum efficiencies for 6.4~keV and 14.4~keV are saturated at the $V_{\rm b}$
and the sensor layer is therefore fully depleted at $V_{\rm b}$~$\gtrsim$~50~V.
The ratio with FI is $\sim 20$\% lower than that with BI at the full depletion. 
This is because 22\% of 6.4~keV X-rays from the frontside is absorbed 
by the circuit and BOX layers with the thickness of $9.2\mu$m in total. 
The full depletion voltage can be translated into a resistivity of 9--10~k$\Omega$$\cdot$cm, which is comparable to the value of the wafer used for the sensor layer.

\begin{figure}[h]
\centering
\includegraphics[width=7.8cm]{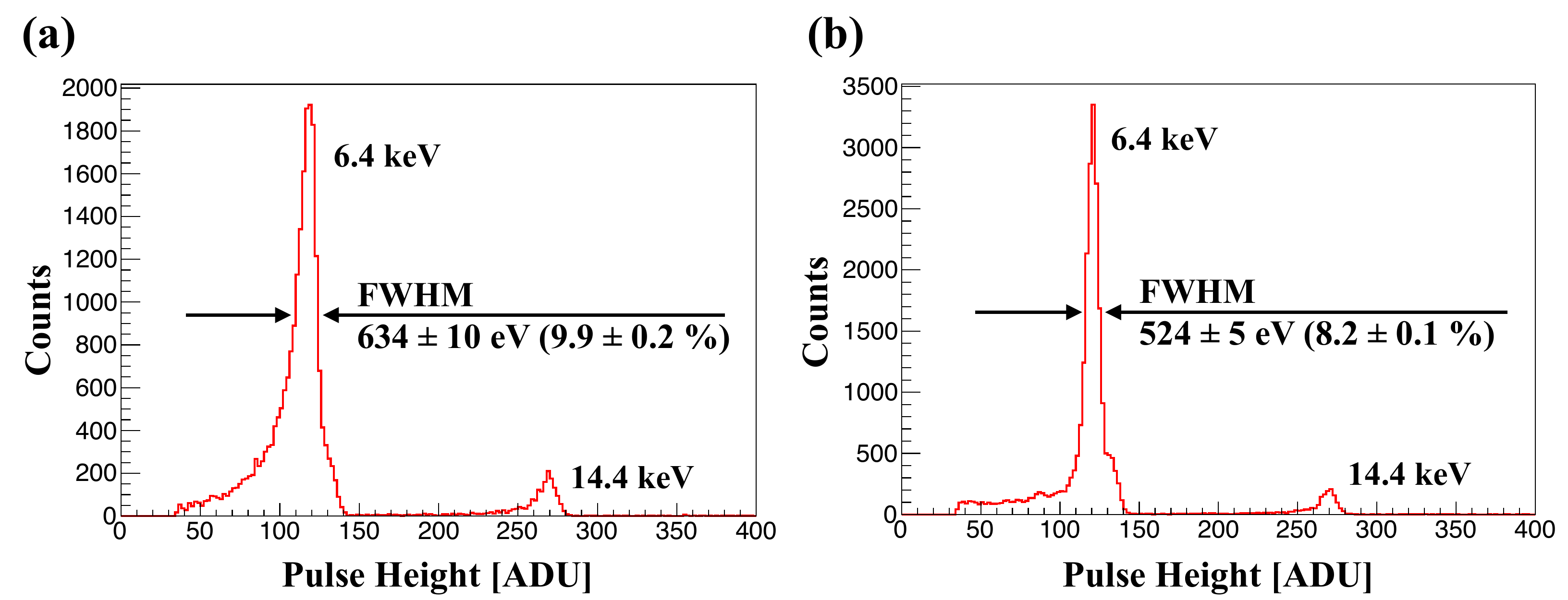}
\caption{(a) X-ray spectrum of a $^{57}$Co radioisotope obtained with the central 32 $\times$ 32 pixels of XRPIX5b with backside illumination.
(b) Same as (a) but with frontside illumination.
The pulse height is shown in analog digital units (ADU). 1~ADU is 488~$\mu$V.
Note that the pulse heights in the figure are amplified by the second stage amplifier with a gain of1.8.}
\label{fig:Spectrum}
\end{figure}

We examined the spectral performance with BI and FI in the frame readout mode,
in which we read out fixed 32~$\times$~32 pixels in each frame after every 1~ms exposure.
We applied $V_{\rm b} = 200$~V to the sensor layer in order to reduce charge diffusion and prevent the charge loss in the sensor layer as much as possible. 
Fig.~\ref{fig:Spectrum} (a) and (b) shows $^{57}$Co spectra obtained with the central 32~$\times$~32 pixels with BI (a) and FI (b) using the identical device.
We use only single-pixel events which are the events when signal charge is collected within one pixel. 
The readout noise and the chip output gain are $\sim$~46~e$^-$~(rms) and $\sim$~18.5~$\mu$V/e$^-$, respectively, 
both with BI and FI.
The leakage current is 3~e$^{-}$~msec$^{-1}$~pixel$^{-1}$.   
The energy resolutions at 6.4~keV are 634~$\pm$~10~eV and 524~$\pm$~5~eV (FWHM) with BI and FI, respectively.
The spectral performance with FI is almost comparable to XRPIX3b \cite{A.Takeda2015}.
However, a low-energy tail degrades the spectral performance in the case of BI.
We speculate that the charge collection efficiency is lower at the backside of the sensor. 
We will examine the details of the electric fields in the sensor layer using a simulation.

\section{Gain uniformity}
\label{Gain uniformity}

\subsection{Uniformity of the whole area of XRPIX5b}
\label{Uniformity of the whole area of XRPIX5b}

\begin{figure}[h]
\centering
\includegraphics[width=6cm]{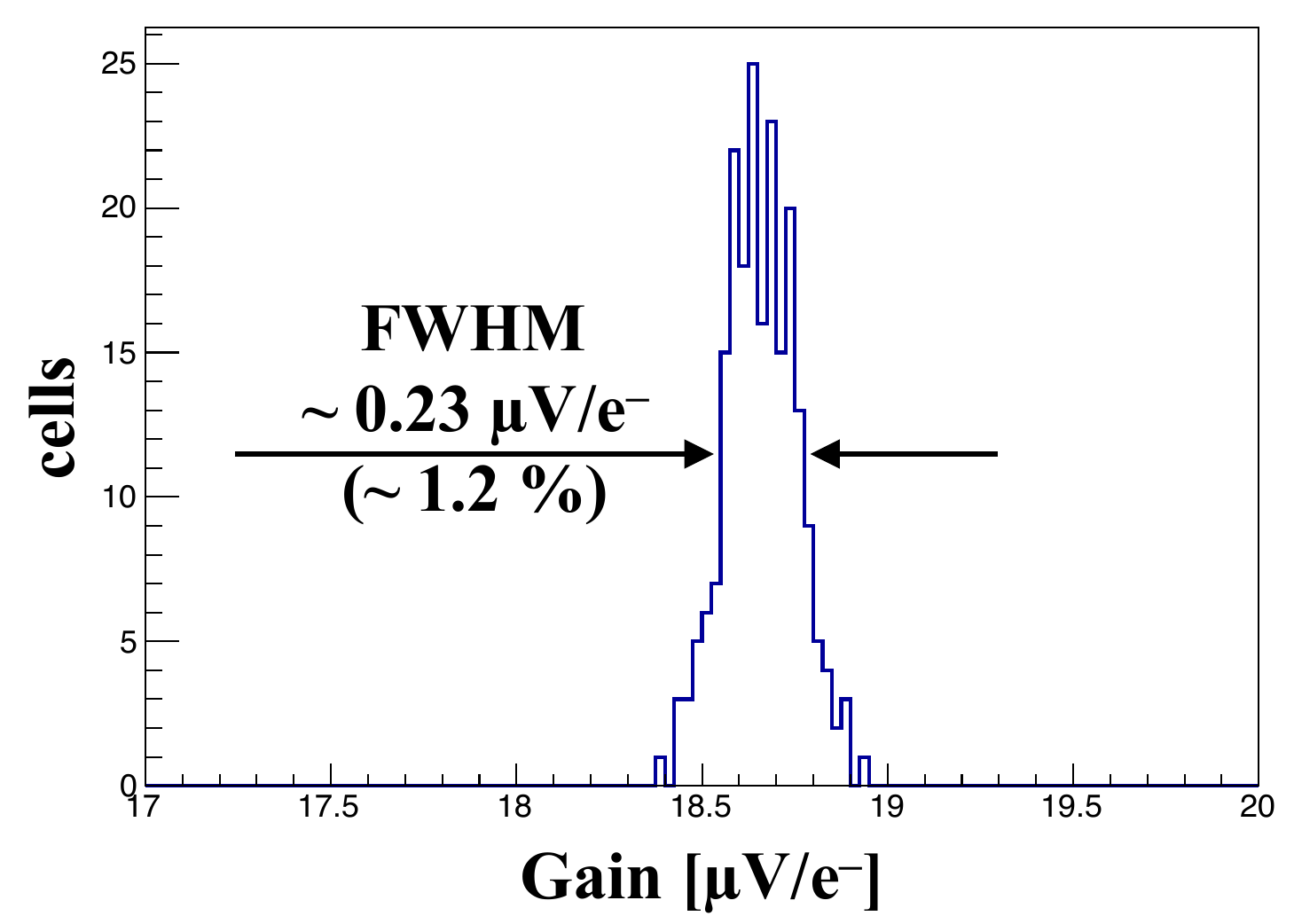}
\caption{Histogram of gains of pixels obtained with XRPIX5b.}
\label{fig:GainHisto_all-cells}
\end{figure}

\begin{figure}[h]
\centering
\includegraphics[width=5cm]{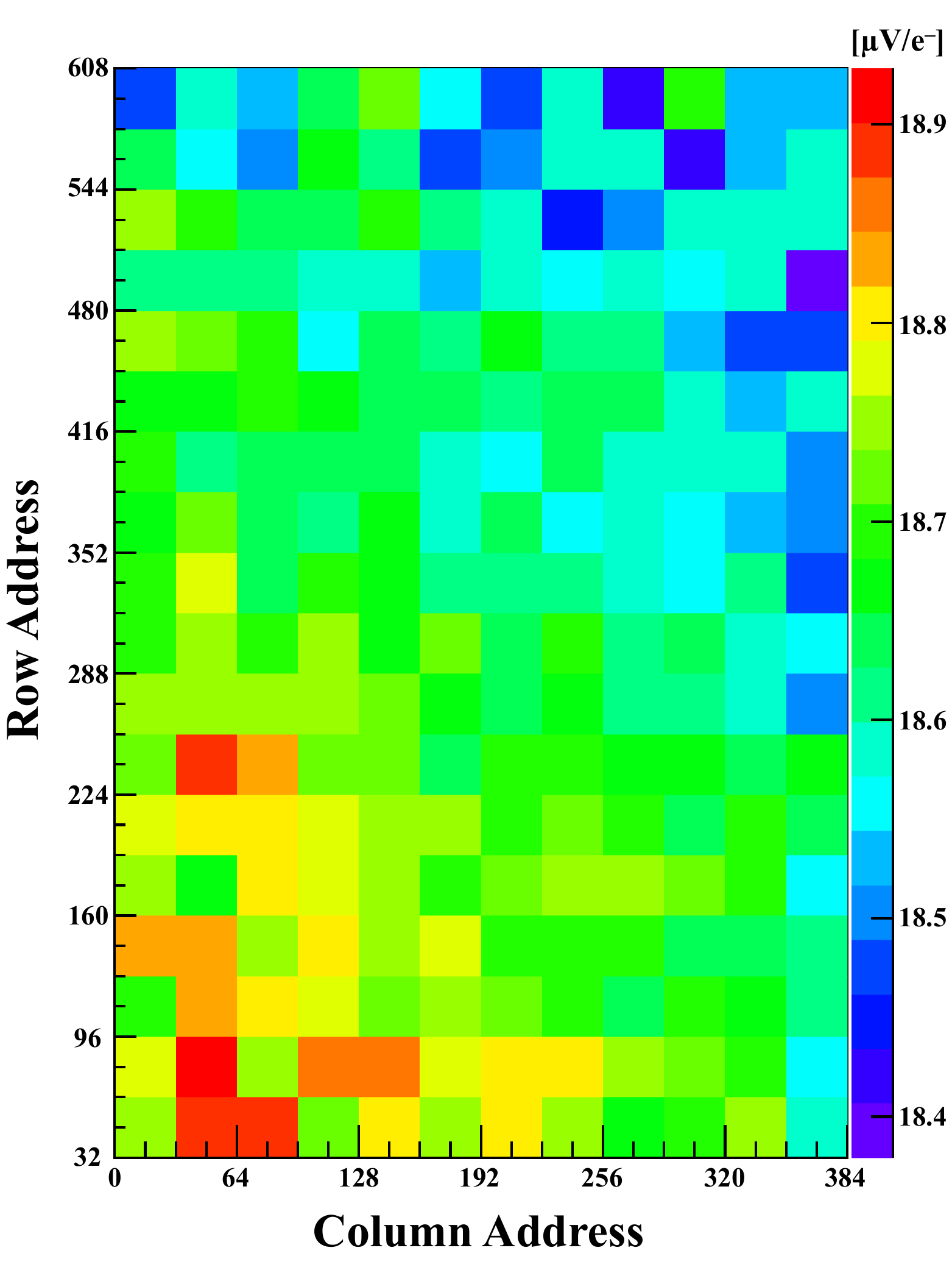}
\caption{Map of gains of pixels obtained with XRPIX5b.}
\label{fig:GainMap_all-cells}
\end{figure}

We successfully obtained X-ray spectra from all the pixels in the frame readout mode, and examined the gain uniformity of XRPIX5b.
Figs.~\ref{fig:GainHisto_all-cells} and \ref{fig:GainMap_all-cells} show a histogram and a map of the gains of the pixels at each location of the sensor.
The gains shown here are obtained by fitting $^{57}$Co spectra with BI of single-pixel events from 32 $\times$ 32 pixels, which we refer to as a cell hereafter.
The data acquisition conditions are the same as those in Section \ref{Spectral performance}.
The statistical errors of  the gains  obtained by fitting spectra are negligible ($<$~0.1~$\mu$V/e$^{-}$) to the gain variation.
Fig.~\ref{fig:GainHisto_all-cells} indicates that the cell-to-cell gain variation is at a 1.2\%, which is not problematic for most of our applications.
However, the gain variation has a systematic tendency.
Fig.~\ref{fig:GainMap_all-cells} shows that the cells away from the analog output pad, which is at the bottom-left corner of XRPIX5b, have lower gains.
The possible cause would be voltage drops due to trace resistance of the signal lines.

\subsection{Uniformity in a cell}
\label{Uniformity in the cells}

We investigated the pixel-to-pixel gain variation in a cell.
Fig.~\ref{fig:GainHisto_[32,64]} shows a histogram of gains of the cell which locates at [column address, row address] = [48, 80].
The other cells have the same tendency as Fig.~\ref{fig:GainHisto_[32,64]}.
The gain variation in a cell is $\sim$~2\%~(FWHM), which probably reflects the gain variation of the amplifiers in each pixel.
The gain variation has little effect on the energy resolution of 8--10\% (Fig.~\ref{fig:Spectrum}).

\begin{figure}[h]
\centering
\includegraphics[width=6cm]{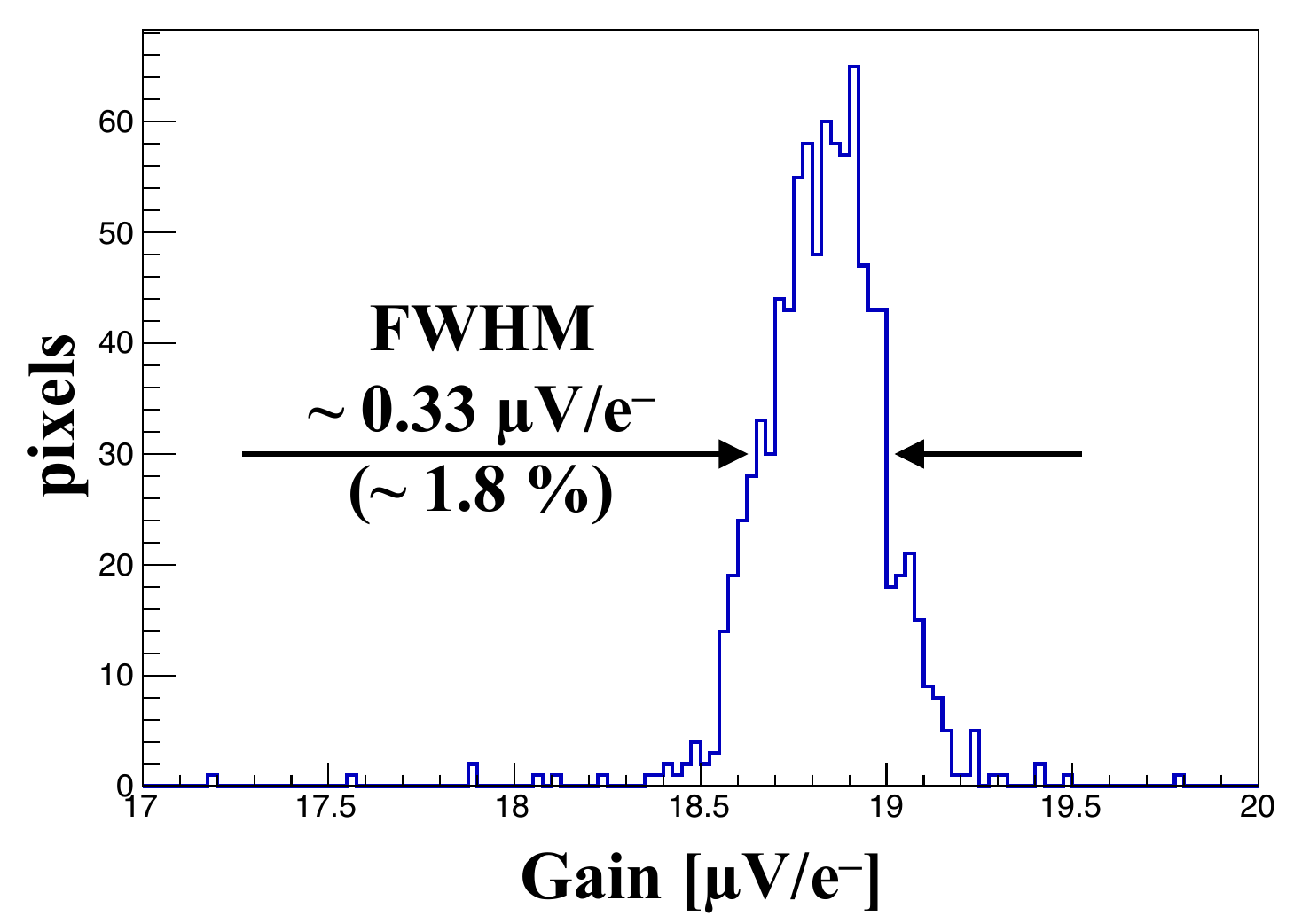}
\caption{Histogram of pixel-to-pixel gains in a cell.}
\label{fig:GainHisto_[32,64]}
\end{figure}

\section{Event-driven readout}
\label{Event-driven readout}

\begin{figure}[h]
\centering
\includegraphics[width=7.5cm]{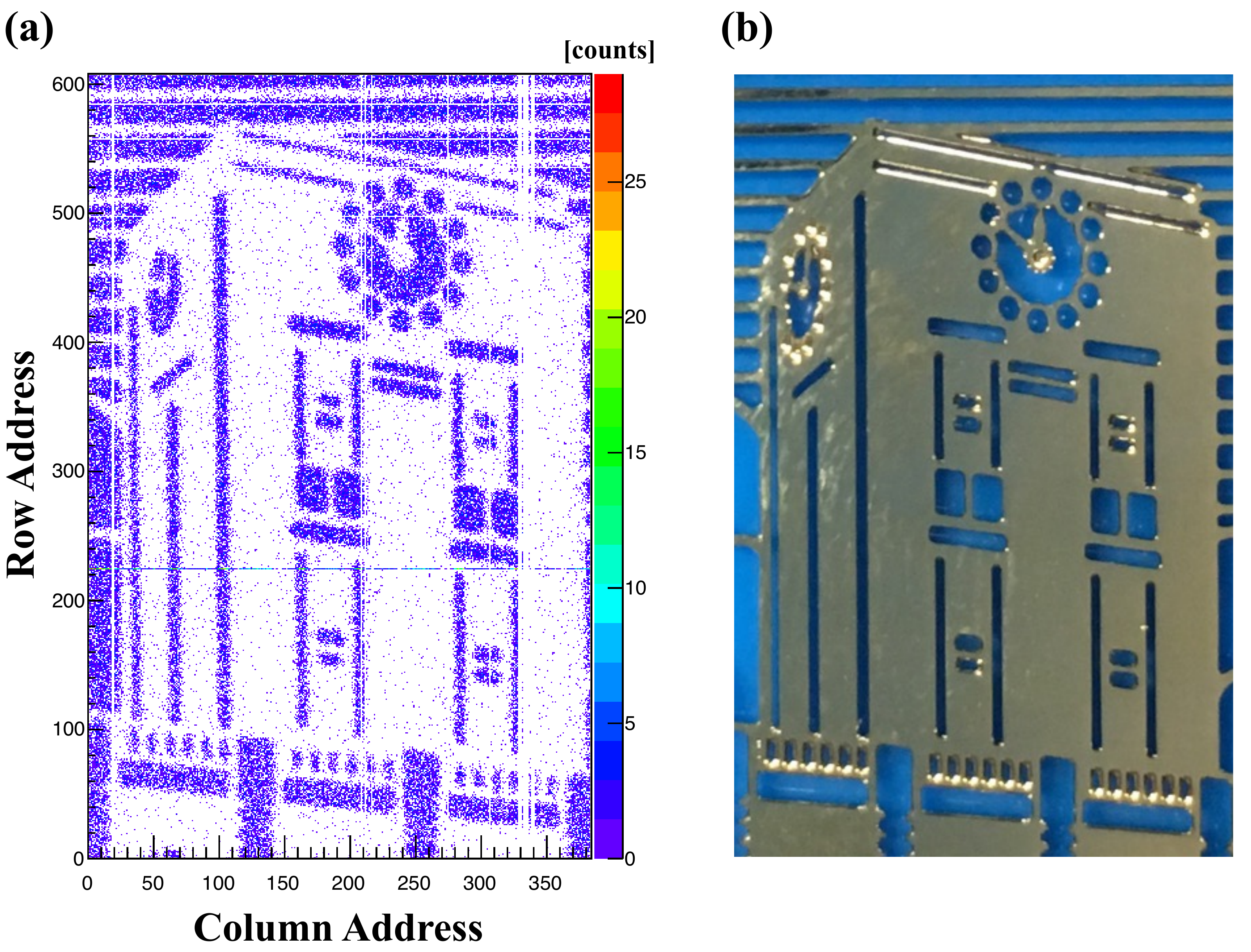}
\caption{(a) X-ray image obtained with XRPIX5b in the event-driven readout mode. (b) Photograph of the metal mask used for the image of (a).}
\label{fig:Image_EventDriven}
\end{figure}

We evaluate the event-driven readout in XRPIX5b in this section.
The details of the readout sequence in the event-driven readout mode are presented in Takeda et al. (2013) \cite{A.Takeda2013},
but there is one different point that the hit address encoder circuit is newly equipped with XRPIX5b.
This circuit allows us to know hit address faster through the parallel digital signal line than through the serial digital signal line used by in previous chips.
Fig.~\ref{fig:Image_EventDriven} shows an X-ray image taken with XRPIX5b in the event-driven readout mode.
We irradiated the detector with 22~keV and 25~keV X-rays from $^{109}$Cd at room temperature. 
We operated it with $V_{\rm b} = 10$~V in this experiment in order to reduce the leakage current.
We confirmed that the trigger circuits in all the pixels operate correctly and X-ray hit addresses are properly output.
While the event rate in Fig.~\ref{fig:Image_EventDriven} is $\sim$~60~Hz per sensor, 
we confirmed that XRPIX5b successfully detects X-rays from $\rm ^{241}$Am at the event rate of $\sim$~570~Hz per sensor without using the mask, 
at the conditions of the backside illumination, $V_{\rm b} = 10$~V and the operating temperature of $-60$~${}^\circ \mathrm{C}$. 
Since the event rate was limited by the irradiated X-ray flux, XRPIX5b would tolerate an even higher event rate.

\begin{figure}[h]
\centering
\includegraphics[width=7.8cm]{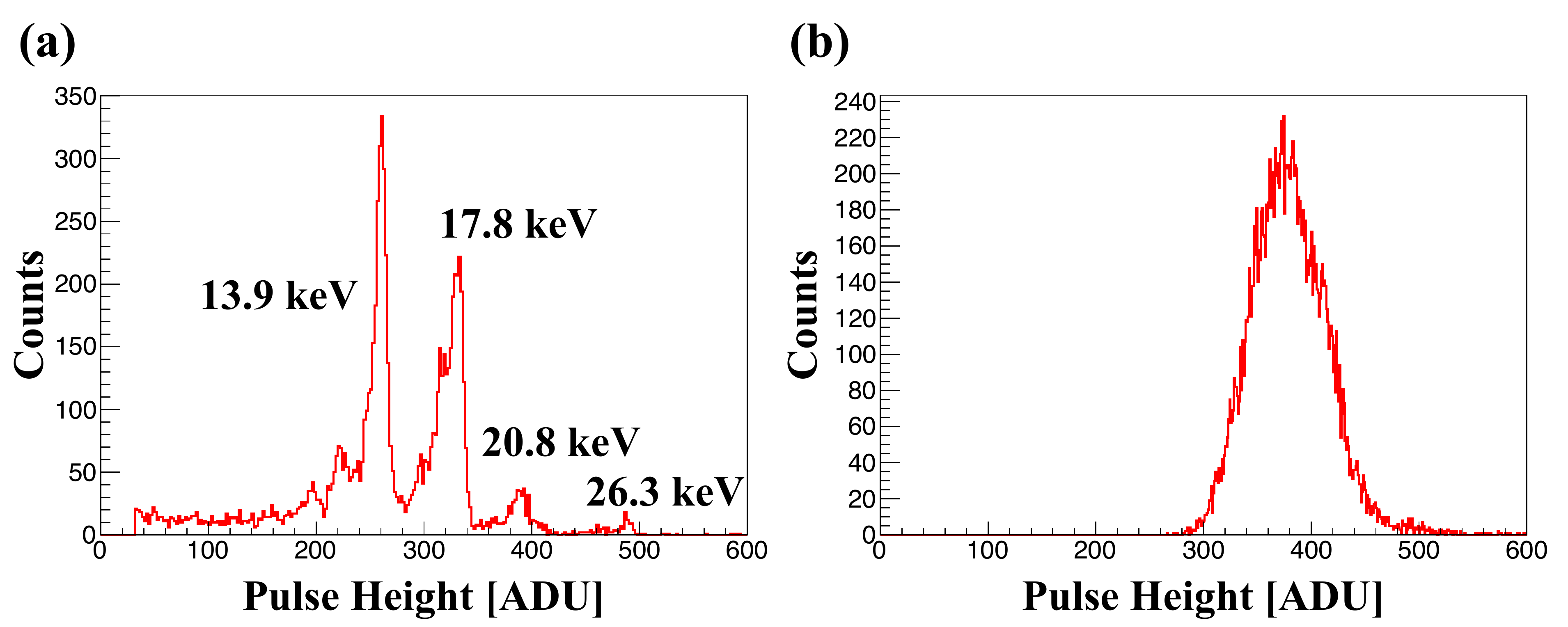}
\caption{(a) X-ray spectrum of a $^{241}$Am radioisotope in the frame readout mode. (b) Same as (a) but in the event-driven readout mode.}
\label{fig:241Am_spec_fr_ed}
\end{figure}

We found a significant problem with X-ray spectrum of XRPIX5b in the event-driven readout mode.
Fig.~\ref{fig:241Am_spec_fr_ed} shows $^{241}$Am spectra obtained in the frame readout mode (a) and in the event-driven readout mode (b). 
The data acquisition conditions are the same as those in Section \ref{Spectral performance}.
The energy resolution in the event-driven readout mode is much worse than that in the frame readout mode.
Takeda~et~al. (2014) \cite{A.Takeda2014} reported that spectral performance in the event-driven readout mode is degraded compared to the frame readout mode
due to the capacitive coupling between the BPW directly connected to the sense node and the trigger digital signal line in the circuit layer.

Ohmura et~al. (2016) \cite{S.Ohmura} reported that the additional silicon layer acts as an electrostatic shield and 
successfully reduces the capacitive coupling between the BPW and the trigger digital signal line in the circuit layer.
Ohmura~et~al. (2016) \cite{S.Ohmura} and Miyoshi~et~al. (2017) \cite{Miyoshi+17} applied the Double SOI (DSOI) structure 
to realize the shield layer. 
Takeda~et~al. in prep. \cite{A.Takeda+18} will report successful reduction of the capacitive coupling 
and improvement of the spectral performance in the event-driven readout mode by using the small size XRPIXs with DSOI, XRPIX6D. 
We are newly processing ``XRPIX7'' with the DSOI structure and with the same imaging area as that of XRPIX5b. 
We expect the DSOI structure will drastically improve the spectral performance of XRPIX7 
in the event-driven readout mode as compared with that of XRPIX5b.

\section{Summary}
\label{Summary}

We processed XRPIX5b, which has the largest imaging area of 21.9~mm $\times$ 13.8~mm, and evaluated its performance.
We successfully obtained X-ray spectra from all the pixels.
The readout noise and the gain are $\sim$~46~e$^-$~(rms) and $\sim$~18.5 $\mu$V/e$^-$, respectively,
XRPIX5b outputs analog signals with little performance variation among the pixels.
The trigger circuits in all the pixels operate correctly even though there is a problem with X-ray spectrum in the event-driven readout mode.

\section*{Acknowledgement}
We acknowledge the valuable advice and great work by the personnel of LAPIS Semiconductor Co., Ltd. 
This study was supported by the Japan Society for the Promotion of Science (JSPS) 
KAKENHI Grant-in-Aid for Scientific Research on Innovative Areas 25109002 (Y.A.), 25109003 (S.K.) 
and 25109004 (T.G.T. \& T.T.), Grant-in-Aid for Scientific Research (B) 23340047 (T.G.T.), 
Challenging Exploratory Research 26610047 (T.G.T.), 
Grant-in-Aid for Young Scientists (B) 15K17648 (A.T.) and Grant-in-Aid for JSPS Fellows 15J01842 (H.M.). 
This study was also supported by the VLSI Design and Education Center (VDEC), 
the University of Tokyo in collaboration with Cadence Design Systems, Inc., and Mentor Graphics, Inc.

\section*{Refference}




\begin{thebibliography}{00}
\bibitem{G.P.Garmire}
G. P. Garmire et al., Advanced CCD imaging spectrometer (ACIS) instrument on the Chandra X-Ray Observatory, 
SPIE, 4851 (2003) 28.
\bibitem{L.Struder}
L. Str\"uder et al., The European Photon Imaging Camera on XMM-Newton: The pn-CCD camera, 
Astronomy \& Astrophysics, 365 (2001) L18.
\bibitem{M.J.L.Turner}
M. J. L. Turner et al., The European Photon Imaging Camera on XMM-Newton: The MOS cameras, 
Astronomy \& Astrophysics, 365 (2001) L27.
\bibitem{K.Koyama}
K. Koyama et al., X-Ray Imaging Spectrometers (XIS) on Board Suzaku, 
Publication of the Astronomical Society of Japan, 59 (2007) S22.
\bibitem{N.Meidinger}
N. Meidinger et al., Report on the eROSITA camera system, 
SPIE, 9914 (2014) 91441W.
\bibitem{H.Tsunemi}
H. Tsunemi et al., Soft x-ray imager (SXI) on-board ASTRO-H, 
SPIE, 9905 (2016) 990535.
\bibitem{T.Tanaka}
T. Tanaka et al., The soft X-ray imager (SXI) aboard the Hitomi satellite, 
J. Astron. Telesc. Instrum. Syst. 4 (2018) 011211.
\bibitem{T.G.Tsuru}
T. G. Tsuru et al., Development and Performance of Kyoto's X-ray Astronomical SOI pixel (SOIPIX) sensor, 
SPIE, 9144 (2014) 914412. 
\bibitem{T.G.Tsuru2018}
T. G. Tsuru et al., Kyoto's event-driven x-ray astronomy SOI pixel sensor for the FORCE mission, 
SPIE, 10709 (2018) 107090H. 
\bibitem{Y.Arai}
Y. Arai et al., Development of SOI Pixel Process Technology, 
Nuclear Instruments and Methods in Physics Research Section A, 636 (2011) S31.
\bibitem{A.Takeda2013}
A. Takeda et al., Design and Evaluation of an SOI Pixel Sensor for Trigger-Driven X-ray Readout,
IEEE Transaction on Nuclear Science, 60 (2013) 586.
\bibitem{A.Takeda2015}
A. Takeda et al., Improvement of spectroscopic performance using a charge-sensitive amplifier circuit for an X-ray astronomical SOI pixel detector,
Journal of Instrumentation, 10 (2015) C06005.
\bibitem{Ryu}
S. G. Ryu et al., First Performance Evaluation of an X-Ray SOI Pixel Sensor for Imaging Spectroscopy and Intra-Pixel Trigger, 
IEEE Transactions on Nuclear Science, 58 (2011) 2528.
\bibitem{Nakashima2012}
S. Nakashima et al., et al. 2012, 
Progress in development of monolithic active pixel detector for X-rayastronomy with SOI CMOS technology,
Physics Procedia, 37, 1373.
\bibitem{A.Takeda2014}
A. Takeda et al., Development and Evaluation of an Event-Driven SOI Pixel Detector for X-Ray Astronomy, 
Proceedings of Science, TIPP2014 (2014) 138.
\bibitem{S.Ohmura}
S. Ohmura et al., Reduction of cross-talks between circuit and sensor layer in the Kyoto's X-ray astronomy SOI pixel sensors with Double-SOI wafer,
Nuclear Instruments and Methods in Physics Research Section A, 831 (2016) 61.
\bibitem{Miyoshi+17}
T. Miyoshi, Y. Arai, Y. Fujita et al., 
Front-end electronics of double SOI X-ray imaging sensors, 
Journal of Instrumentation, 12 (2017) C02004
\bibitem{A.Takeda+18}
A. Takeda et al., in prep. 


\end{thebibliography}


\end{document}